\documentclass[aps,prr,twocolumn,superscriptaddress,showpacs,amsmath,amssymb, amsfonts,10pt]{revtex4-1}

\usepackage{amsmath}
\usepackage{amsfonts}
\usepackage{amssymb}
\usepackage{graphicx}
\usepackage{color}
\usepackage{bm}
\usepackage[normalem]{ulem}
\usepackage[colorlinks,bookmarks=false,urlcolor=blue, citecolor=blue, linkcolor = blue]{hyperref}
\usepackage{comment}

\usepackage{lipsum}

\newcommand{\oldtext}[1]{\textcolor{red}{\sout{#1}}}

\newcommand{\rr}{\mathbf{r}}

\newcommand{\dd}{\mathrm{d}}

\newcommand{\qq}{\mathbf{q}}

\begin{document}

\title{
Emergent Kardar-Parisi-Zhang phase in quadratically driven condensates
}

\author{Oriana K. Diessel}
\affiliation{Max-Planck-Institute of Quantum Optics, Hans-Kopfermann-Strasse 1 , 85748 Garching, Germany}

\author{Sebastian Diehl}
\affiliation{Institute for Theoretical Physics, University of Cologne, Z\"{u}lpicher Strasse 77, 50937 Cologne, Germany}

\author{Alessio Chiocchetta}
\affiliation{Institute for Theoretical Physics, University of Cologne, Z\"{u}lpicher Strasse 77, 50937 Cologne, Germany}

\begin{abstract}
In bosonic gases at thermal equilibrium, an external quadratic drive can induce a Bose-Einstein condensation described by the Ising transition, as a consequence of the explicitly broken U(1) phase rotation symmetry down to $\mathbb{Z}_2$.
However, in physical realizations such as exciton-polaritons and nonlinear photonic lattices, thermal equilibrium is lost and the state is rather determined by a balance between losses and external drive. A fundamental question is then how nonequilibrium fluctuations affect this transition.
Here, we show that in a two-dimensional driven-dissipative Bose system the Ising phase is suppressed and replaced by a nonequilibrium phase featuring Kardar-Parisi-Zhang (KPZ) physics. Its emergence is rooted in a U(1)-symmetry restoration mechanism enabled by the strong fluctuations in reduced dimensionality. Moreover, we show that the presence of the quadratic drive term enhances the visibility of the KPZ scaling, compared to two-dimensional U(1)-symmetric gases, where it has remained so far elusive. 
\end{abstract}

\date{\today}

\maketitle

How the absence of thermal equilibrium affects the properties of matter is one of the fundamental questions of many-body physics, with far-reaching consequences in the engineering of novel materials, the development of quantum technologies, and the understanding of active and living matter.  
In nonequilibrium systems, the lack of detailed balance can radically modify the collective behaviours typical of equilibrium systems.  
Accordingly, novel phases can be expected, such as non-reciprocal (or chiral) phases in active matter \cite{Saha2020,fruchart2020}, quantum optical platforms~\cite{Hanai2019,Young2020} and ultracold atoms~\cite{dogra2019}, or dissipative time crystals in many-body quantum systems~\cite{Russomanno2017,Iemini2018,buvca2019,Lazarides2020}.

An intriguing aspect concerns the impact of nonequilibrium fluctuations in low spatial dimensions. At equilibrium, the Mermin-Wagner theorem forbids the spontaneous breaking of a continuous symmetry in spatial dimensions $d \leq 2$ for systems with short-ranged interactions. 
Out of equilibrium, the theorem does not hold: two-dimensional flocks~\cite{Toner1995} or driven quantum spin chains~\cite{Prosen2008,Prosen2010} can feature transitions to phases with long-range order. On the converse, it was shown that the Berezinskii–Kosterlitz–Thouless (BKT) phase transition, expected for equilibrium Bose gases in two spatial dimensions, is erased in their driven-dissipative counterpart and replaced by a disordered phase featuring a Kardar-Parisi-Zhang (KPZ) scaling of the phase fluctuations~\cite{Altman2015}. 
A promising candidate to experimentally observe this scaling are exciton-polaritons fluids in microcavities~\cite{Kasprzak2006,CarusottoRMP2013}, although the length scales at which its signatures are expected are dramatically larger than the typical system sizes~\cite{Roumpos12,Nitsche14,Dagvadorj2015,Caputo2018}.

The fate of nonequilibrium systems with discrete symmetries is less explored. At equilibrium, the absence of Goldstone modes dismiss them from the scope of the Mermin-Wagner theorem and, accordingly, they can exhibit order also in two dimensions. This is the case for the arguably most paradigmatic phase transition, namely the Ising transition. Among its many incarnations, the Ising phase transition can be realized in bosonic gases in the presence of 
an externally imprinted pair creation term: in ultracold atoms, this can be induced by coupling to a molecular condensate~\cite{radzihovsky2008} (see also Ref.~\cite{Jiang2011} for a wire of fermionic atoms), by using a parametric down conversion scheme in microcavities~\cite{Carusotto1999,Bardeen2012}, or by Feshbach-like resonances in polariton-biexcitons ~\cite{Carusotto2010,Takemura2014} or Rydberg polaritons~\cite{Alberton21}.
At equilibrium, the bosons undergo a Bose-Einstein condensation (BEC) transition belonging to the Ising universality class~\cite{radzihovsky2004,Romans2004}.  
In optical systems, the unavoidable presence of incoherent processes causes a departure from equilibrium. Still, recent numerical analyses showed that these driven-dissipative models can undergo a BEC transition characterized by either the quantum or classical Ising universality class~\cite{Savona2017,Rota2019,Verstraelen2019,Verstraelen2020}.

In this paper, we show that the absence of thermal equilibrium suppresses the Ising phase transition in a two-dimensional, driven-dissipative Bose gas, in favour of an emerging KPZ phase. 
Our two main results are summarized as follows.
First, we find that the long-wavelength description of the quadratically-driven Bose gas is given by a driven sine-Gordon equation for the phase degree of freedom. In two spatial dimensions, this dynamics is dominated by the KPZ scaling at long wavelengths, ultimately resulting in the suppression of the BKT and Ising phases, present instead at equilibrium. 
%
%
Second, we find that the presence of the quadratic drive reduces the scale at which the KPZ physics sets in, enhancing its visibility in finite-size systems. 
This hold promises for identifying this physics in two spatial dimensions, where experimental realizations remain so far elusive~\cite{Halpin2014,Takeuchi2018}.

\emph{Microscopic model}--- We consider a gas of quadratically-driven and dissipative bosons, whose dynamics is described by the master equation
\begin{equation}
    \partial_t \hat{\rho} = -i[\hat{H},\hat{\rho}] + \int_\rr \sum_n \left[ \hat{L}_n \rho \hat{L}_n^\dagger- \frac{1}{2}\{\hat{\rho}, \hat{L}_n^\dagger \hat{L}_n \} \right],
\end{equation}
with $\hat{\rho}$ the system's density matrix, $\hat{H}$ the Hamiltonian and $\hat{L}_n = \hat{L}_n(\rr)$ Lindblad operators. The quadratic drive can be regarded as a process coherently creating or destroying two particles at a given position. The Hamiltonian is thus given by
\begin{equation}
\label{eq:Hamiltonian}
\!\!\!\hat{H} \!= \!\!\int_\rr \!\left[ \frac{\nabla \hat{\psi}^\dagger\nabla \hat{\psi}}{2m}  + \delta \hat{\psi}^\dagger \hat{\psi} +\frac{G}{2}(\hat{\psi}^2+ \hat{\psi}^{\dagger 2}) + \frac{U}{2} \hat{\psi}^{\dagger2}\hat{\psi}^2 \right],
\end{equation}
%
with $m$ the mass of the bosons, $\delta > 0$ the detuning between the bosonic fundamental frequency and the drive frequency, 
and $U> 0$ the particle interaction. The quadratic drive comes with a strength $G$, and we can set $G >0$  without loss of generality, by absorbing its phase into a redefinition of the fields. 
%
The presence of further incoherent processes, such as single particle losses and pump, as well as two-particle losses, is included via the Lindblad operators $\hat{L}_\text{1l} = \hat{\psi}$, $\hat{L}_\text{1p} = \hat{\psi}^\dagger$, and $\hat{L}_\text{2l} = \hat{\psi}^2$, respectively. In the following, we will assume the single-particle pump to be weaker than single-particle losses.

Since we are interested in the critical properties of this model, we neglect quantum fluctuations, as they are irrelevant compared to the statistical fluctuations induced by the incoherent processes~\cite{Chiocchetta2014,sieberer2016keldysh}. This approximation allows us to treat $\hat{\psi}$ as a stochastic field rather than an operator: its dynamics is accordingly described by the Langevin equation
\begin{equation}
\partial_t \psi= -\big(-K \nabla^2+r + u|\psi|^2 \big)\psi - i G\psi^*+\zeta,
\label{eq:Langevin}
\end{equation}
with $K, r, u$ complex numbers, and $\zeta$ a Gaussian, zero-average white noise with correlations $\langle \zeta(\rr,t)\zeta^*(\rr',t')\rangle = 2\sigma \delta(t-t')\delta^{(2)}(\rr-\rr')$. 
The imaginary parts of $K, r, u$ (in the following denoted by a ``$c$'' subscript) correspond to coherent couplings describing reversible dynamics, while their real parts (in the following denoted by a ``$d$'' subscript) correspond to dissipative couplings 
representing irreversible processes.
Moreover, Eq.~\eqref{eq:Langevin} includes terms which, while zero at the microscopic level, are expected to be generated by coarse-graining, e.g., $K_d$, describing spatial diffusion.

For $G=0$, Eq.~\eqref{eq:Langevin} is invariant under the U(1) transformation $\psi \to e^{i\alpha} \psi, \psi^* \to e^{-i\alpha} \psi^* $, and it is known as complex Ginzburg-Landau equation~\cite{Cross1993,Aranson2002}, or as driven-dissipative Gross-Pitaevski equation in the context of exciton-polaritons~\cite{CarusottoRMP2013}.
For finite values of $G$, Eq.~\eqref{eq:Langevin} is invariant under the $\mathbb{Z}_2$ transformation $\psi\to -\psi$, $\psi^* \to -\psi^*$, and it is known as periodically-driven complex Gross-Pitaevski equation ~\cite{Aranson2002}.

\emph{Driven sine-Gordon equation} ---
A simple mean-field analysis of Eq.~\eqref{eq:Langevin} shows that a phase transition is expected for $G>G_c$, predicting the spontaneous breaking of the $\mathbb{Z}_2$ symmetry and the emergence of a condensate. This result is expected to be qualitatively robust in higher spatial dimensions $d>2$, while in lower dimensions fluctuations can dramatically modify the mean-field result.

In order to assess the effect of fluctuations, we proceed in the spirit of the hydrodynamic theory for quasicondensates~\cite{Popov1972,Altman2015}, and we represent the bosonic complex field as $\psi(\rr,t) = \chi(\rr,t)e^{i\theta(\rr,t)}$, with $\chi$ and $\theta$ real fields associated with density and phase fluctuations. By assuming that a condensate exists, with a density determined by the saddle-point equations, the dynamics is dominated by configurations of $\chi$ around that value. The density field $\chi$ is gapped and can  therefore be  eliminated adiabatically from the dynamics (see App.~\ref{app:mapping}). This results in the following effective equation for the phase
\begin{equation}
\label{eq:Langevin_phase}
\eta\,  \partial_t \theta=\gamma \nabla^2 \theta -2 g \sin(2 \theta)+\frac{\lambda}{2}(\nabla \theta)^{2} + F + \xi,
\end{equation}
with $\xi$ a zero-average Gaussian white noise with correlations $\left\langle \xi(\mathbf{r}, t) \xi\left(\mathbf{r}^{\prime}, t^{\prime}\right)\right\rangle=2 D \delta^{(2)}(\mathbf{r}-\mathbf{r}^{\prime}) \delta(t-t^{\prime})$.
The microscopic values of the six parameters $\eta,\gamma, g, \lambda, F, D$ are given by 
%
%
\begin{align}
    \eta &= 1, \quad\quad\quad\qquad\qquad\,\,\,\,\, \gamma = K_d+\frac{u_c}{u_d}K_c,\nonumber\\
    g &=\frac{G}{2}\sqrt{1+\frac{u_c^2}{u_d^2}}, \quad\qquad\,\,
    \lambda =2\left(-K_c+\frac{u_c}{u_d}K_d\right),\\
     F &=-r_c+\frac{u_c}{u_d}r_d, \quad\qquad \label{eq:conversion}
    D = \frac{\sigma}{2\chi_0^2}\left(1+\frac{u_c^2}{u_d^2}\right). \nonumber
\end{align}
The $\mathbb{Z}_2$ symmetry of Eq.~\eqref{eq:Langevin} is inherited by
Eq.~\eqref{eq:Langevin_phase} as an invariance under the transformation $\theta \to \theta + m\pi $, for all odd integers $m$.
The properties of the phase $\theta$ derived from the solutions of Eq.~\eqref{eq:Langevin_phase} can be directly translated into the correlations of the original complex fields $\psi, \psi^*$ via 
\begin{subequations}
\label{eq:psi_to_theta}
\begin{align}
     \langle \psi(\rr) \rangle & \approx \chi_0 e^{i\theta_0}\, e^{-\frac{1}{2}\langle \theta(\rr)^2\rangle},
     \\
    \langle \psi(\rr) \psi^*(0)\rangle &\approx \chi_0^2\,  e^{\langle \theta(\rr)\theta(0) \rangle - \langle \theta(\rr)^2 \rangle},
\end{align}
\end{subequations}
with $\theta_0$ the saddle point value of $\theta$. The previous relations are obtained by neglecting the fluctuations of $\chi$, and retaining only the leading terms in the cumulant expansion of $\langle e^{i (\theta(\rr)-\theta(0))} \rangle$.
\begin{figure*}[t!]
	\centering

	\includegraphics[width=\linewidth]{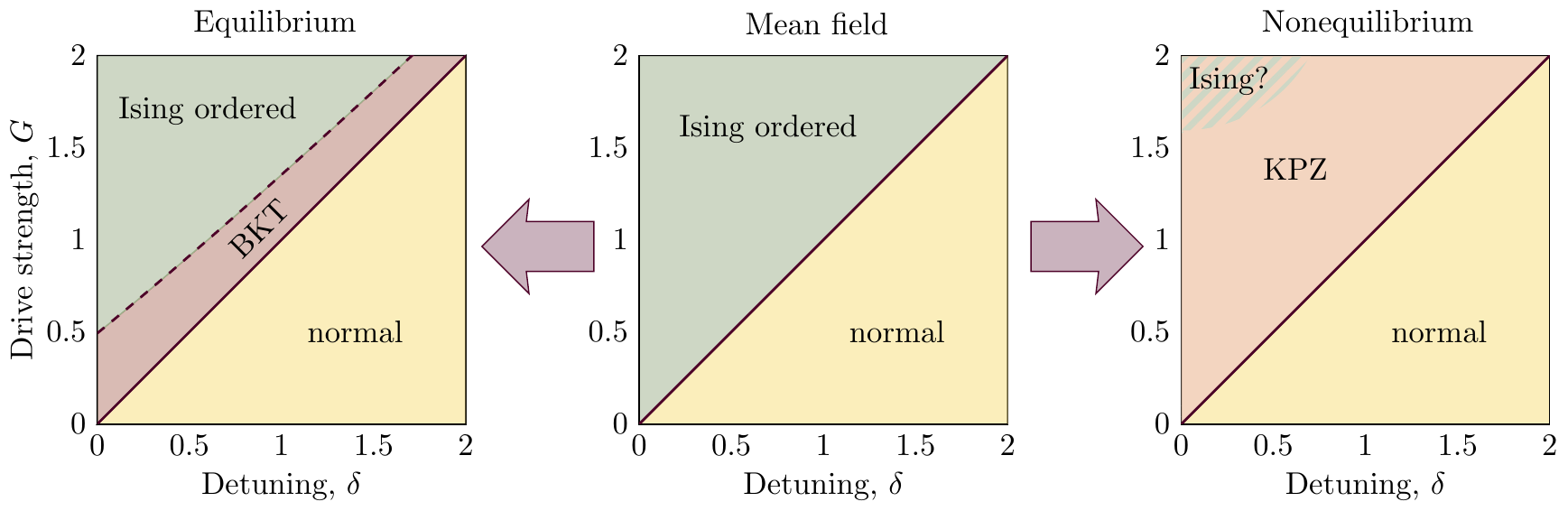}
	\caption{Effect of fluctuations on the mean-field phase diagram of Eqs.~\eqref{eq:Langevin} and~\eqref{eq:Langevin_phase}, in- and out-of-equilibrium. \emph{Center}--- Mean-field phase diagram of a quadratically driven, open condensate, as a function of the imprinted pairing strength $G$ and the detuning $\delta$. 
	\emph{Left}--- Equilibrium phase diagram (see App. \ref{app:PD_derivation} for the derivation). Fluctuations give rise to an additional intermediate phase featuring BKT scaling.
	\emph{Right}---
	Nonequilibrium phase diagram of the full model~\eqref{eq:Langevin_phase}. The Ising phase is replaced by a phase which features KPZ scaling.
	A residual Ising phase (denoted by the striped region) may persist in the non-perturbative regime of large $G$, inaccessible to our method.
	\label{fig:PhaseDiagram}}
\end{figure*}
%

A first insight into the solution of Eq.~\eqref{eq:Langevin_phase} can be gained by considering two limiting cases and only then the general scenario: 

\emph{(i) KPZ limit} --- For $g=0$, the equation possesses a U(1) symmetry, realized by the invariance under the transformation $\theta \to \theta + \alpha$, with $\alpha$ any real number, and the drift term $F$ can be removed by a gauge transformation $\theta \to \theta + F t/\eta$. Equation~\eqref{eq:Langevin_phase} thus reduces to the pristine KPZ equation~\cite{KPZ_86}.
In two spatial dimensions, the massless, KPZ-like fluctuations of the phase were shown to erase the BKT phase usually expected in equilibrium Bose gases, and replace it with a disordered phase~\cite{Altman2015}.

\emph{(ii) Equilibrium limit} --- Another relevant limiting case is given by thermal equilibrium. This is achieved when the condition $K_c/K_d = r_c/r_d = u_c/u_d$ is satisfied~\cite{Sieberer2013,Altman2015}, which entails the validity of the fluctuation-dissipation theorem, or, more generally, the presence of the associated thermal symmetry of the Keldysh action~\cite{Sieberer2015}. In this case, one has $\lambda=0$ and $F=0$, and Eq.~\eqref{eq:Langevin_phase} reduces to the relaxational dynamics of a sine-Gordon field, whose renormalization was first studied in relation to the roughening transition of crystal surfaces~\cite{nozieres1987}. 

This model predicts two different phases, depending on the relevance of the sine term. 
In the first phase, the field $\theta$ is massive, which is signalled by the coupling $g$ being relevant in the RG sense. As a consequence, the value of $\langle \theta(\rr)^2\rangle$ is infrared-convergent, while $\langle \theta(\rr)\theta(0) \rangle$ decays exponentially at long distances. Accordingly, Eqs.~\eqref{eq:psi_to_theta} predict the order parameter $\langle \psi \rangle$ to be finite and long-range order is established, indicating that the system lies in the ordered phase with a spontaneously broken $\mathbb{Z}_2$ symmetry.
In the second phase, $g$ is irrelevant in the RG sense, and  $\theta$ becomes massless. Accordingly, $\langle \theta(\rr)^2\rangle$ is infinitely large as a consequence of the infrared divergence, while $\langle \theta(\rr)\theta(0) \rangle - \langle \theta(\rr)^2 \rangle$ grows logarithmically, implying an algebraic decay of $\langle \psi^*(\rr) \psi(0)\rangle $.
This then suggest that long-range order is no longer supported, and the condensed phase is replaced by a BKT phase characterized by quasi-long-range order. This is the usual case for two-dimensional Bose gases with U(1) symmetry (i.e., $G=0$ in Eq.~\eqref{eq:Hamiltonian}). 

Summarizing, for an equilibrium gas in two dimensions, the following three phases are expected: a normal fluid with short-range correlations (corresponding to the mean-field solution without condensate), a BKT phase with quasi-long-range order, and a $\mathbb{Z}_2$-symmetry-broken phase with long-range order. The corresponding phase diagram in terms of $G$ and $\delta$ is reported in Fig.~\ref{fig:PhaseDiagram}, (see App.~\ref{app:PD_derivation} for derivation). Analogous phases have been obtained for the ANNNI model~\cite{selke1988annni,suzuki2012quantum} and the XYZ spin chain in transverse field~\cite{Sela2011,pinheiro2013x}, which share the same effective dimensionality and $\mathbb{Z}_2$ symmetry with the present model. For these spin chains, the correct hydrodynamic description is provided by  the Luttinger liquid theory with sine-Gordon perturbations.    

\emph{(iii) Full problem }--- In the full Eq.~\eqref{eq:Langevin_phase}, the KPZ fluctuations wash out the sine-Gordon physics, thus destabilizing the phases predicted at thermal equilibrium. The renormalization analysis of this equation was first performed in Refs.~\cite{Hwa1991, Rost1994} in order to study the effect of nonlinearities on the roughening transition of crystal surfaces. There, it was shown that the KPZ physics dominates over large distances. We will show that this has dramatic implications for driven-dissipative Bose gases, as the equilibrium ordered and BKT phases are destabilized by nonequilibrium fluctuations, and replaced by a phase with short-range order, see Fig.~\ref{fig:PhaseDiagram}. 

\emph{Absence of long-range order}---
The long-wavelength physics of Eq.~\eqref{eq:Langevin_phase} can be conveniently studied using a perturbative renormalization group approach. The idea consists in treating $g$ and $\lambda$ as perturbations around the Gaussian model, and in deriving an effective long-wavelength theory by progressively integrating out high-energy modes. The form of the couplings of the long-wavelength model is then encoded in a set of flow equations.
We will consider two different RG schemes, derived in Refs.~\cite{Rost1994} and~\cite{Ettouhami2003}, respectively, and discussed in App.~\ref{app:RG_derivation}. The equations are expressed in terms of the dimensionless quantities $\bar{g} \equiv g/\Lambda^2$, and $\bar{F} \equiv F / \Lambda^2$.
Before proceeding to a more detailed analysis, we discuss the qualitative behaviour of the solution of the RG equations.

If the system is in thermal equilibrium,  then the equations reduce to the ones for the relaxational sine-Gordon model of Ref.~\cite{nozieres1987}. 
If $\bar{g}=0$, instead, the equations reduce to the usual ones for KPZ~\cite{tauber2014critical}: the noise level $D$ and the effective temperature $T$ flow to infinity, indicating the relevance of the KPZ scaling. 
Finally, if both $\bar{g}_0$ and $\lambda_0$ are finite, the KPZ nonlinearity $\lambda$ dominates over the sine-Gordon one $\bar{g}$, which eventually renormalizes to zero.
\begin{figure}[t!]
	\centering
	\includegraphics[width=1.0\linewidth]{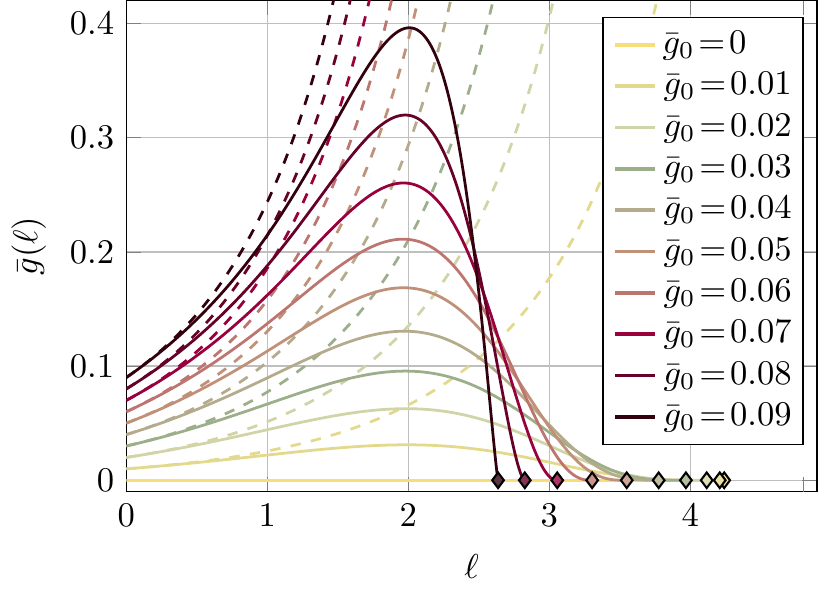}
	\caption{Flow of $\bar{g}(\ell)$ for different initial values of $\bar{g}_0$. 
	Parameters for the solid curves: $\gamma_0=0.3,\, T_0=1,\, \lambda_0=0.4,\, \eta_0=1, \bar{F}_0=0$. Parameters for the dashed curves: $\gamma_0=0.3,\, T_0=1,\, \lambda_0=0.0,\, \eta_0=1, \bar{F}_0=0$. 
	The diamond symbols denote the onset of a divergence in the RG flow, occurring at the values of $\ell^*$ reported in Fig.~\ref{fig:l_star}.
	\label{fig:different_U}}
\end{figure}
Typical flows of $\bar{g}$ are shown in Fig.~\ref{fig:different_U}: for $\lambda=0$ and $\bar{F}=0$,  $\bar{g}(\ell)$ grows indefinitely (dashed curves), signalling that the field $\theta$ is in the gapped phase. For finite initial values of $\lambda_0$ or $\bar{F}_0$, however, the growth of $\bar{g}$ is interrupted, and it flows back to zero, indicating the irrelevance of the sine-Gordon term. The diamond symbols denote the onset of a divergence in the RG flow (see below).

At long wavelengths, the phase correlations are then expected to be captured by the KPZ exponents, i.e., $\langle \theta(\rr)\theta(0)\rangle- \langle \theta(\rr)^2\rangle \sim -|\rr|^{2\chi} $, with $\chi\approx 0.38$~\cite{Halpin-Healy2012}. Moreover, the value of $\langle \theta(\rr)^2\rangle$ diverges due to long-wavelength fluctuations. Accordingly, by replacing these values in Eqs.~\eqref{eq:psi_to_theta}, we find that complex fields are short-range correlated via a stretched exponential, leading to the conclusion that no phase transition can take place.
Whether the ordered phase is completely removed, or survives for large values of the two-particle drive (corresponding to large values of $g$) cannot be determined from our analysis, as the RG analysis is not valid for non-perturbative values of $g$. 
Finally, here we neglected the presence of topological excitations, such as vortices and anti-vortices, which are essential to describe the transition between the KPZ and a normal, featureless phase~\cite{Wachtel2016,Sieberer2016,Sieberer2018,Zamora2020,Gladilin2019,Gladilin2020}. The impact of the $\mathbb{Z}_2$ symmetry on these excitations is left for future work. 

\emph{Enhancement of KPZ physics}---
%
%
An essential question concerns the visibility of the predicted 2D KPZ physics in experimental systems or numerical simulations with limited size.
%
It turns out that the length scale $L^*$ above which the KPZ physics becomes visible is usually very large, and can exceed the accessible systems' size: this is the case for, e.g., the roughening transition in crystal surfaces~\cite{Balibar1992}, and for exciton-polaritons in two-dimensional microcavities~\cite{Altman2015,Dagvadorj2015,Zamora2017,Comaron2018,Mei2020}.
Here we show that the presence of a sine-Gordon nonlinearity can actually lower the length scale $L^*$, thus enhancing the visiblity of the 2D KPZ.

\begin{figure}[t!]
	\centering
\includegraphics[width=1.0\linewidth]{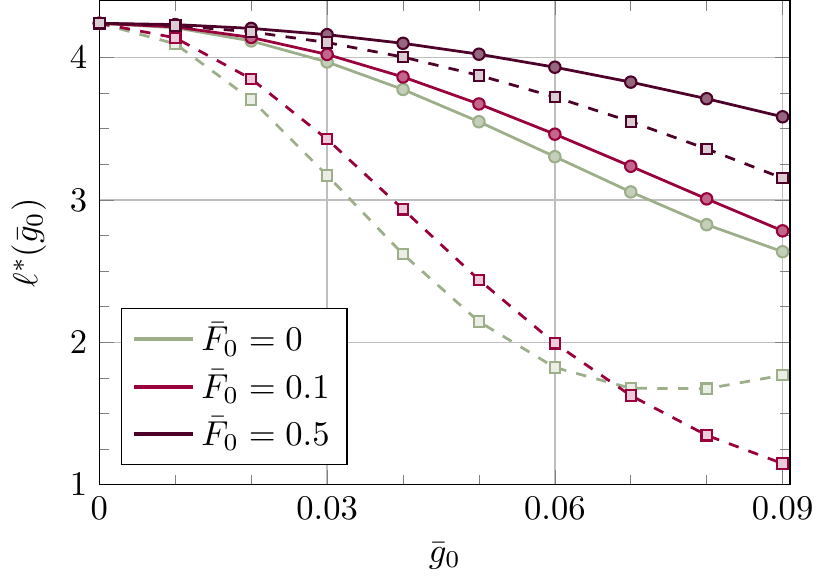}
	\caption{ RG scale for the KPZ crossover as a function of the microscopic sine-Gordon nonlinearity $\bar{g}_0$, for different values of $\bar{F}_0$. The solid and dashed lines correspond to the RG schemes derived in Ref.~\cite{Rost1994} and~\cite{Ettouhami2003}, respectively (cf. App.~\ref{app:RG_derivation}). 
	Parameters: $\gamma_0=0.3,\, T_0=1,\, \lambda_0=0.4,\, \eta_0=1$.
	\label{fig:l_star}
	}
\end{figure}

The value of $L^*$ can be extracted from the solution of the flow equations~\cite{Natterman1992}.
To illustrate this, it is convenient to first focus on the pure KPZ case of Eqs.~\eqref{eq:RG_equations}, i.e., $\bar{g}=0$. In this case, the relevant RG equation is the one for the effective temperature $T$ in Eq.~\eqref{eq:RG_T} with $\gamma$ and $\lambda$ constant under the RG flow. 
%
$T(\ell)$ features a divergence for finite values of the flow parameter $\ell$, namely $\ell^* = 8\pi \gamma^3/(T_0 \lambda^2)$, with $T_0$ the initial value of $T$. The value of $\ell^*$ determines therefore the physical length scale above which the KPZ scaling is visible via $L^* = \xi_0 e^{\ell^*}$, with $\xi_0$ some microscopic length scale. As $L^*$ is exponentially sensitive to the value of $\ell^*$, finding conditions to minimize $\ell^*$ is crucial to observe the KPZ physics. For finite values of $\bar{g}$, the value of $\ell^*$ cannot be determined analytically, but it can be extracted from the divergence of the numerical solutions.  We computed $\ell^*$ for different values of $\bar{g}_0$ and $\bar{F}_0$: the results are reported in Fig.~\ref{fig:l_star}.
Since $\ell^*$ is not a universal quantity, we extracted its value using two different RG schemes (cf. App.~\ref{app:RG_derivation}), finding the same qualitative behavior. 

Our results indicate that the value of $\ell^*$ generically decreases as a function of $\bar{g}_0$. The decrease can be optimized by varying the value of $\bar{F}$, which, corresponding to the laser detuning (cf. Eq.~\eqref{eq:conversion}), is an experimentally tunable parameter. 
The value of $\ell^*$ can be reduced by up to a factor $4$ upon reaching $\bar{g}_0 \sim 0.1$, indicating that $L^*$ can be reduced by four orders of magnitude compared to the case with $\bar{g}_0=0$. This result implies a dramatic improvement of the visibility of the KPZ scaling in two dimensional driven-dissipative gases, where it has so far remained elusive. As an example, in exciton-polariton fluids in the optical parametric oscillator regime, the KPZ length scale was predicted to be $\sim 10^3 \mu m$ in the bad-cavity regime~\cite{Zamora2017}, which is one order of magnitude larger than the typical size in current experiments~\cite{Exp1,Exp2,Exp3}. The presence of a quadratic drive would then bring the KPZ length scale well below the system size, unveiling the corresponding scaling.

\emph{Outlook}--- We showed that, in two-dimensional quadratically-driven Bose gases, the absence of thermal equilibrium leads to an emerging phase characterized by KPZ scaling. Correspondingly, the BKT and Ising phases expected at thermal equilibrium are suppressed. Moreover, we discovered that the presence of a quadratic drive may shrink the length scale at which the KPZ physics occurs, thus enhancing its visibility in systems with finite size.
Our results open novel perspectives for the detection of nonequilibrium phases of matter in experimental platforms, in particular exciton-polaritons in microcavities and nonlinear photonic lattices. There, a quadratic drive can serve as a tool to enhance the nonequilibrium nature of driven-dissipative condensates, and may provide the necessary assist to experimentally access the unexplored physics of the 2D KPZ equation.

\emph{Acknowledgments} --- We acknowledge support by the funding from the European Research Council (ERC) under the Horizon 2020 research and innovation programme, grant agreement No. 647434 (DOQS) and by the DFG (CRC 1238 project number 277146847 - project C04). O.~K.~D. is supported by a fellowship of the International Max Planck Research School for Quantum Science and Technology (IMPRS-QST).

\appendix

\section{Derivation of driven sine-Gordon equation}
\label{app:mapping}
We provide here a more detailed discussion of the mapping used to derive Eq.~\eqref{eq:Langevin_phase}.
Starting from Eq.~\eqref{eq:Langevin}, we insert the phase amplitude representation for the field 
$\psi(\rr,t) = \chi(\rr,t)e^{i\theta(\rr,t)}$, and separate real and imaginary parts, obtaining the two equations:
%
\begin{subequations}
\begin{align}
    \partial_t \theta & = -r_c + K_c \left[ \frac{\nabla^2 \chi}{\chi}  - (\nabla \theta)^2\right] + K_d \left[2\frac{\nabla \chi}{\chi} \nabla\theta +\nabla^2\theta\right]\nonumber \\ 
    &  -G\sin(2\theta) + \text{Im}\left[\frac{\zeta e^{-i\theta}}{\chi}\right], \\
    \partial_t \chi & = -r_c \chi K_d\left[\nabla^2 \chi -\chi(\nabla\theta)^2 \right] - K_c\left[2\nabla\chi\nabla \theta + \chi\nabla^2\theta \right] \nonumber \\
    & -u_d \chi^3 - G\chi \sin(2\theta) + \text{Re}\left[\zeta e^{-i\theta}\right].
\end{align}
\end{subequations}
By linearizing the equation for $\chi$ around its saddle point value $\chi_0$, the gapped nature of the fluctuations becomes evident. Assuming that these fluctuations are small compared to $\chi_0$, we neglect spatial and time derivatives of $\chi$ from the previous equations, and we can adiabatically eliminate $\chi$ from the remaining equations. By further performing the shift $\theta \to \theta + \theta_0 $, with $\tan(2\theta_0) = u_d/u_c$, we obtain the effective equation for $\theta$ given in Eq.~\eqref{eq:Langevin_phase}.

\section{Derivation of equilibrium phase diagram}
\label{app:PD_derivation}

In this appendix, we derive the equilibrium phase diagram in Fig.~\ref{fig:PhaseDiagram} in the main text.
To this end, we consider the purely relaxational dynamics given by
\begin{equation}
\partial_t \psi=-\frac{\delta H}{\delta \psi^*}+\zeta,
\end{equation}
with $H$ the Hamiltonian given in Eq~\eqref{eq:Hamiltonian}. Then, we perform a mean field analysis by taking the expectation value of the previous equation, and using the space- and time-independent Ansatz $\langle\psi(\mathbf{x},t)\rangle=\psi_0=\chi_0 e^{i\theta_0}$. The resulting equation predicts, for $\delta>0$, two different values for the amplitude $\chi_0$, namely an ordered phase with a finite expectation value of $\psi$ for $G\geq\delta$, and a disordered phase with a vanishing field expectation value for $G<\delta$.

Next, we investigate how fluctuations affect the ordered phase predicted by the mean field. As the amplitude field is gapped, the relevant low-energy excitations are the fluctuations of the phase field, whose dynamics is described by 
\begin{equation}
\eta\,  \partial_t \theta=\gamma \nabla^2 \theta -2 g \sin(2 \theta)+\xi,
\end{equation}
where we identify $\gamma=1/2m$ and $2g=G$. The BKT RG flow of these two parameters 
shows two basins of attraction,  whose separatrix can be approximated by a line $g=\beta(\gamma-\gamma_c)$, which has a zero at $\gamma_c=D/2\pi$ with $D=\sigma/2\chi_0^2$, and a slope of $\beta\approx -1.7$. The region where $g$ is irrelevant, and therefore the phase with long range order is replaced by a BKT phase (cf. discussion in the main text) is described by
\begin{equation}
    \frac{G}{2U}<\beta\left(\frac{\sigma}{4\pi}\frac{1}{G-\delta}-\frac{1}{2mU}\right). 
\end{equation}
In Fig.~\ref{fig:PhaseDiagram}, we show the phase diagram for the values $\sigma=4$ and $m=U=1$.

\section{RG equations}
\label{app:RG_derivation}

We discuss here the RG equations used, corresponding to the schemes used in Refs.~\cite{Rost1994} and~\cite{Ettouhami2003}, respectively.
Both schemes are based on partitioning the phase fluctuations $\theta$ and the noise $\xi$ (cf. Eq.~\eqref{eq:Langevin_phase}) into fast and slow modes, the first corresponding to modes with momenta $\qq$ lying in the shell $\Lambda (1-\dd \ell) |\qq| \leq \Lambda $, and the seconds to modes $|\qq|<\Lambda(1-\dd \ell)$. Here $\Lambda$ is the ultraviolet cutoff of the model, while $\dd \ell$ corresponds to an infinitesimal dimensionless number controlling the width of the momentum shell. The fast modes are then integrated out, generating an effective action for the slow modes, which is computed perturbatively in $g$ and $\lambda$. Finally, the momenta, frequencies, and fields are rescaled in order to restore the original cut off $\Lambda$: the resulting action provides the flow of the coupling constants upon taking the limit $\dd \ell \to 0$. 

\begin{figure}[t!]
	\centering
	\includegraphics[width=\linewidth]{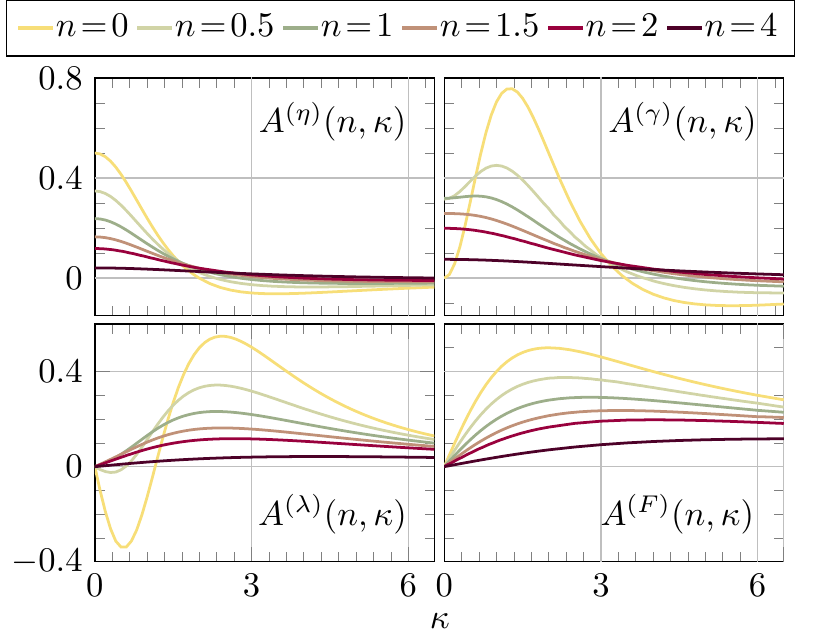}
	\caption{
	Functions defined in Eq.~\eqref{eq:A_functions_Rost}, as function of $\kappa$ and for different values of $n$.
	\label{fig:AllA}}
\end{figure}
This program can be carried out in different ways. In Ref.~\cite{Rost1994}, the perturbative corrections are evaluated at level of the Langevin function~\eqref{eq:Langevin_phase}, using the Nozieres-Gallet scheme~\cite{nozieres1987}. In Ref.~\cite{Ettouhami2003}, instead, Eq.~\eqref{eq:Langevin_phase} 
is represented as Martin-Siggia-Rose-Janssen-De Dominicis functional~\cite{tauber2014critical}, and then the perturbative corrections are computed in a fashion similar to the usual sine-Gordon renormalization (see, e.g.,~Ref.\cite{gogolin2004bosonization}). The two schemes lead expectedly to two different schemes of RG equations, which lead to quantitatively different RG flows. However, the fixed-points structure is the same, as a consequence of universality. 
The equations can be brought in the general form:
\begin{subequations}
\label{eq:RG_equations}
\begin{align}
\frac{\dd \bar{g}}{\dd \ell} & =\left(2-\frac{T}{\pi \gamma}\right) \bar{g}, \\
\frac{\dd \gamma}{\dd \ell} & =\frac{2 T }{\pi \gamma^2 }  A^{(\gamma)}(n,\kappa) \bar{g}^{2} , \\
\frac{\dd \eta}{\dd \ell} & =\frac{8 T \eta}{\pi\gamma^3}  A^{(\eta)}(n,\kappa) \bar{g}^{2} , \\
\frac{\dd T}{\dd \ell}& =\frac{T^2 \lambda^2 }{8\pi \gamma^3}+ \frac{8T^2  }{ \pi \gamma^{3} } A^{(T)}(n,\kappa) \bar{g}^2, \label{eq:RG_T} \\
\frac{\dd \lambda}{\dd\ell} & =\frac{8T}{\pi \gamma^2 }   A^{(\lambda)}(n,\kappa) \bar{g}^{2}, \\
\frac{\dd \bar{F}}{\dd \ell}& =2 \bar{F}+\frac{T \lambda}{4\pi \gamma}-\frac{4T}{\pi\gamma^2}  A^{(F)}(n,\kappa) \bar{g}^{2} ,
\end{align}
\end{subequations}
with $T\equiv D/\eta$ the effective temperature, $\bar{g} \equiv g/\Lambda^2$, $\bar{F} \equiv F/\Lambda^2$, $\kappa \equiv 2\bar{F}/\gamma$, and $n\equiv T/\pi \gamma$. The functions $A(n,\kappa)$ take different values depending on the renormalization scheme. The scheme followed in Ref.~\cite{Ettouhami2003} leads to functions independent of $n$, which read:
\begin{subequations}
\label{eq:A_functions_Keldysh}
\begin{align}
   A^{(\eta)}(\kappa)&=2\,\frac{4-\kappa^2}{(4+\kappa^{2})^2},\\
   A^{(\gamma)}(\kappa)&=2\,\frac{32-12 \kappa^{2}-\kappa^{4}}{(4+\kappa^{2})^{3}},\\
   A^{(T)}(\kappa)&= \frac{4\kappa^{2}}{(4+\kappa^{2})^{2}},\\
   A^{(\lambda)}(\kappa)&=8\,\frac{\kappa^{3}+20\kappa}{(4+\kappa^{2})^{3}},\\
   A^{(F)}(\kappa)&=\frac{2\kappa}{4+\kappa^{2}}.
\end{align}
\end{subequations}

The functions obtained in Ref.~\cite{Rost1994} read, instead:

\begin{subequations}
\label{eq:A_functions_Rost}
\begin{align}
   A^{(\eta)}(n,\kappa)&=\int_{0}^{\infty}\!\! \dd x \dd {\rho} \, \rho^3  g(x,\rho; n) \cos (\kappa x {\rho}^{2}), \\
   A^{(\gamma)}(n,\kappa)&=\int_{0}^{\infty}\!\! \dd x \dd {\rho} \,\frac{\rho^3}{x}  g(x,\rho; n) \cos (\kappa x {\rho}^{2}),\\
   A^{(\lambda)}(n,\kappa)&=\int_{0}^{\infty}\!\! \dd x \dd {\rho} \,\frac{\rho^3}{x}  g(x,\rho; n) \sin (\kappa x {\rho}^{2}),\\
   A^{(F)}(n,\kappa)&=\int_{0}^{\infty}\!\! \dd x \dd {\rho} \, \frac{\rho}{x}  \, g(x,\rho; n)\sin (\kappa x {\rho}^{2}), 
\end{align}
\end{subequations}
with $A^{(T)}(n,\kappa)=0$ and 
\begin{equation}
g(x,\rho; n) \equiv J_{0}({\rho})  e^{-\frac{1}{4 x}-x {\rho}^{2}-2 n \varphi({\rho}, x)},
\end{equation}
with
\begin{equation}
\label{eq:phi_function}
\varphi({\rho}, x)=\int_0^{1}\!\frac{\dd k}{k}\left(1-J_0(k{\rho})e^{-k^2x\rho^2}\right).    
\end{equation}
The form of the functions $A$ is shown in Fig.~\ref{fig:AllA} as a function of $\kappa$ and for different values of $n$ (cf. also Refs.~\cite{nozieres1987,Rost1994}). The numerical evaluation of the functions $A^{(\eta)},A^{(\gamma)},A^{(\lambda)}$ and $A^{(F)}$ is a computationally demanding task, given the double integration in $\rho$ and $x$, and the integration in the function $\varphi(\rho,x)$. This task is simplified for $n \gg 1$ or $\kappa \gg 1$: in those cases only values around $\rho=0$ give significant contribution. Accordingly, by approximating $g(x,\rho;n) \approx e^{-\frac{1}{4 x}- n\frac{\rho}{2}(1+4x)}$, the integral over $\rho$ can be computed exactly in the saddle-point approximation. The functions can then be approximated as:
\begin{subequations}
\begin{align}
  A^{(\eta)}(n,\kappa) & \approx 2 \int_0^{\infty} \!\! \dd x\,   e^{-\frac{1}{x}} \frac{n^2(1+x)^2 -(\kappa x)^2}{[n^2 (1+x)^2 +(\kappa x)^2]^2}, \\
  A^{(\gamma)}(n,\kappa) & \approx \int_0^{\infty} \!\! \dd x\, e^{-\frac{1}{x}} \frac{8}{x} \frac{n^2(1+x)^2 -(4\kappa x)^2}{[n^2 (1+x)^2 +(\kappa x)^2]^2}, \\
  A^{(\lambda)}(n,\kappa) & \approx \int_0^{\infty} \!\! \dd x\, e^{-\frac{1}{x}}  \frac{16 \kappa n (1+x)}{[n^2 (1+x)^2 +(\kappa x)^2]^2}, \\
A^{(F)}(n,\kappa) & \approx \int_0^{\infty} \!\! \dd x\, e^{-\frac{1}{x}}  \frac{2 \kappa }{n^2 (1+x)^2 +(\kappa x)^2}, 
\end{align}
\end{subequations}
which can be easily numerically evaluated.  

\bibliography{bibliography.bib}

\end{document}